\documentclass[12pt]{article}
\usepackage{amsmath,amssymb,latexsym}
\usepackage{epsfig}

\newcommand{\be}{\begin{equation}}
\newcommand{\ee}{\end{equation}}
\newcommand{\bea}{\begin{eqnarray}}
\newcommand{\eea}{\end{eqnarray}}
\newcommand{\beqar}{\begin{eqnarray*}}
\newcommand{\eeqar}{\end{eqnarray*}}

\begin{document}

\newpage
\bigskip

\bigskip
\bigskip
\bigskip

\centerline{\bf \Large First Law, Counterterms }
\medskip
\centerline{\bf \Large and }
\medskip
\centerline{\bf \Large Kerr--AdS$_5$ Black Holes }

\bigskip
\bigskip
\bigskip

\centerline{ Adel M. Awad{$^\sharp$}\footnote{On leave from
Department of Physics, Faculty of Science, Ain Shams University,
Cairo, Egypt.} }
\bigskip
\bigskip
\bigskip

\centerline{\it Department of Physics and Astronomy, University of
Kentucky,} \centerline{\it Lexington, KY 40506, U.S.A.}
%\medskip
\centerline{\it Center for Theoretical Physics, British University
in Egypt,}\centerline{\it  Sherouk City 11837, P.O. Box 43,
Egypt.}
\bigskip

\centerline{\tt $^\sharp$adel@pa.uky.edu}

\bigskip

\begin{abstract}
\medskip We apply the counterterm subtraction technique to calculate the action and other quantities for the
Kerr--AdS black hole in five dimensions using two boundary
metrics; the Einstein universe and rotating Einstein universe with
arbitrary angular velocity. In both cases, the resulting
thermodynamic quantities satisfy the first law of thermodynamics.
We point out that the reason for the violation of the first law in
previous calculations is that the rotating Einstein universe, used
as a boundary metric, was rotating with an angular velocity that
depends on the black hole rotation parameter. Using a new
coordinate system with a boundary metric that has an arbitrary
angular velocity, one can show that the resulting physical
quantities satisfy the first law.

\end{abstract} \newpage \baselineskip=18pt
\setcounter{footnote}{0}

\section{Introduction}

Since the discovery of the AdS/CFT correspondence
\cite{maldacena,witten,gubklebpoly}, there has been considerable
interest in Anti de--Sitter (AdS) spacetimes and their physical
quantities. These quantities can reveal many important properties
of the strongly coupled field theory on the boundary. In the last
few years, there has been a debate concerning the thermodynamical
quantities of Kerr--AdS black holes and the first law of black
hole thermodynamics. This debate started with the work of Gibbons,
Perry and Pope
 \cite{GPP1} re-calculating the thermodynamical quantities of
Kerr-AdS black holes in various dimensions using the background
subtraction technique. Comparing their results with previous
results \cite{klemm1, hawkingtwo, adel3}, they showed that their
quantities obey the first law of thermodynamics, while the
quantities produced by some previous calculations, including those
using counterterm method, do not obey the first law. This gave the
impression that the counterterm technique did not produce the
correct thermodynamical quantities for these Kerr--AdS solutions.
In this article we show that using the standard counterterm
calculation (i.e., without adding any new counterterms) for the
Kerr--AdS$_5$ case, one can produce physical quantities that
satisfy the first law of thermodynamics. Here we take the boundary
metric to be the non-rotating Einstein universe, similar to
\cite{GPP1}. One should notice that the boundary metric chosen
here is not the one used in \cite{adel,adel3}. In this case the
thermodynamic quantities did not seem to satisfy the first law,
i.e., \be dE= TdS+\Omega_i\, dJ^i. \label{1stlaw}\ee These
apparently different results of the counterterm method, naturally
raise the question: Why do some choices of the boundary metric
satisfy the first law and others do not? Let us remember that,
according to the AdS/CFT duality, all boundary metrics in a given
conformal class should produce the same quantities for a specific
AdS solution. Of course, in certain cases, {\it e.g.}, when the
conformal symmetry is anomalous, some quantities such as energy
and action, depend on the chosen boundary metric. But, we known
how these quantities change upon going from one boundary metric to
another in the same conformal class. This should not affect the
validity of the first law. In \cite{skenderis05} Papadimitrious
and Skenderis have formulated a variational problem for AdS
gravity with Dirichlet boundary conditions. Their formulation
naturally reproduces the known counterterms that leave the AdS
action finite. Furthermore, they were able to show that all
asymptotically locally AdS black holes satisfy the first law.
Here, we discuss the particular case of Kerr--AdS$_5$ and ask the
question; What went wrong in choosing the rotating Einstein
universe (REU) as a boundary in \cite{adel,adel3} calculations? We
show that the reason for the violation of the first law is not
that the REU was chosen as boundary metric but that it was
rotating with an angular velocity, $\Omega_{\infty}=-a/l^2$, that
depends on the black hole rotation parameter, $a$. The boundary
angular velocity can be interpreted as that of an observer, or a
rotating gas, at infinity which does not have to dependent on the
black hole parameters. Working with a new coordinate system for
Kerr--AdS$_5$ with arbitrary angular velocity at infinity, one can
show that the relevant physical quantities satisfy the first law.
Another interesting consequence of using the new coordinate
system/boundary is that the first law is satisfied whether we used
the energy associated with $\partial_t$ or
$\partial_t+\Omega_{\infty}\partial_{\phi}$. This leads to the
conclusion that in the counterterm method angular velocities, or
other quantities, associated with a boundary metric should be
independent from the black hole parameters, otherwise, the first
law might be violated. It is interesting to notice that if we
allow the angular velocity to vary this will lead to an additional
term in the first law due to a surface tension on the boundary.
The surface tension is nothing but the Casimir pressure in the
boundary theory. We show that the existence of such a pressure
will not affect the stability of the system since its
compressibility is non-negative.

\section{Counterterms and Gravitational Actions}
The AdS/CFT duality states that
\begin{equation}
<e^{\int\,\phi_0(x)\,{\cal O}(x)}>_{CFT}=Z_{AdS}(\phi)
\end{equation}
where $\phi$ is a bulk field and $\phi_{o}$ is its value on the
boundary. If $\phi=g$ is the metric on AdS and $\phi_o=\gamma$ is
its value on the boundary, then ${\cal O}=T_{\mu\nu}$ is the
energy momentum tensor of the boundary field theory. In the low
energy limit, we have
\begin{equation}
Z_{CFT}(\gamma)\simeq e^{-I_{AdS}(g)}
\end{equation}
i.e., AdS gravitational action acts as the effective CFT action.
The gravity action for asymptotically anti-de-Sitter space ${\cal
M}$, with boundary $\partial {\cal M}$\cite{Gibbons} is given by,
\begin{eqnarray}
I_{\rm bulk}+I_{\rm surf}= -{1 \over 16 \pi G}\int_{\cal M}
d^{n+1}x \sqrt{-g}\left(R+{n(n-1) \over l^2}\right)-{1 \over 8 \pi
G } \int_{\partial {\cal M}} d^{n}x \sqrt{-h} \,K.
\end{eqnarray}
Where, $\Lambda{=}{-n(n{-}1)/2l^2}$ is the cosmological constant
and the second term is the Gibbons--Hawking boundary term.
$h_{ab}$ is the induced metric on the boundary and $K$ is the
trace of the extrinsic curvature $K^{ab}$ of the boundary. Since
asymptotically AdS spacetimes have infinite volumes, this action
diverges unless one uses some regularization method. The most
commonly used regularization techniques are: i) the background
subtraction technique and ii) the counterterm subtraction
technique. The background subtraction technique utilizes the fact
that divergent contributions in the AAdS space action is due to
the asymptotic region (i.e., where $r\rightarrow \infty$).
Therefore, one can obtain a finite action by subtracting the AdS
space action from the AAdS action. The main problem with such a
technique is that any physics common between the two manifolds
cancels out and will not be carried by the resulting finite
action. For example, physical quantities on the gravity side dual
to Casimir energy and conformal anomaly vanish on the gravity side
upon using the background subtraction method. On the other hand if
one calculates such quantities on the field theory side, one
obtains non-vanishing expressions. This creates a clear mismatch
between the two sides of the duality, since this piece of action
carries important information about the strongly coupled CFT on
the boundary. The counterterm subtraction technique uses the fact
that divergent contributions to the AAdS gravitational actions can
be written as surface terms that depend on the metric $h$ and its
covariant derivatives \cite{witten}. By calculating these
expressions and using them as counterterms one can define a finite
gravitational action\cite{Henningson}.
\begin{eqnarray} I_{\rm ct}={1 \over 8 \pi G} \int_{\partial {\cal
M}}d^{n}x\sqrt{-h}\Biggl[ \frac{(n-1)}{ l}-{l{\cal R} \over
2(n-2)}\Bigg]\ . \label{theterms}
\end{eqnarray} Here ${\cal R}$ and ${\cal R}_{ab}$ are the Ricci
scalar and tensor for $h$.  Using these counter-terms one can
construct a divergence--free stress--energy tensor al'a Brown and
York from the finite action $I{=}I_{\rm bulk}{+}I_{\rm
surf}{+}I_{\rm ct}$ by defining (see Ref. \cite{Brown} for more
details):
\begin{eqnarray}  T^{ab}&=& {2 \over \sqrt{-h}} {\delta I
\over \delta h_{ab}}\ . \label{stressone}
\end{eqnarray}
We will be interested in using this stress tensor to calculate
conserved quantities for AdS solutions, specifically the total
energy and angular momentum of kerr-AdS$_5$ solution. The
Brown-York conserved charge is given by \cite{Brown}: \be
Q_{\xi}=\int_{\Sigma}d^{D-2}x\sqrt{\sigma}
u^{\mu}T_{\mu\nu}\xi^{\nu}\ . \label{charges} \ee where $\xi$ is a
 Killing vector and $u_{\mu}{=}-N\,t,_{\mu}$, while $N$ and
$\sigma$ are the lapse function and the space-like metric which
appear in the ADM--like decomposition of the boundary metric \be
ds^2=-N^2 dt^2+\sigma_{ab}(dx^{a}+N^a dt)(dx^b+N^b dt)\ . \ee It
is worth mentioning that the metric restricted to the boundary,
$h_{ab}$, diverges due to the infinite conformal factor that
depends on a radial coordinate that we might call $r$. One can
have a well defined boundary metric $\gamma$ as

\begin{equation}
\gamma_{ab}=\lim_{r\to\infty}{\Omega^2}h_{ab}\ . \label{newmetric}
\end{equation}
where $\Omega$ is some positive function with first order pole in
$r$. This defines a conformal structure on the boundary
\cite{witten} rather than a specific boundary metric {\it i.e.}, a
class of boundary metrics for a specific AdS solution which are
related by conformal transformations. As we stressed in the
introduction, this puts all possible metrics in a given conformal
class on equal footing. In principle, one can use any of them to
calculate the action and conserved quantities of a given AdS
solutions up to pieces dual to CFT conformal anomalies and Casimir
energies which should not affect the thermodynamic properties of
such a solution.

As a consequence of the counterterm subtraction technique one can relate the field theory's energy momentum tensor predicted by the
duality ${\widehat T}^{ab}$ and the CFT energy momentum tensor
\cite{robstress}:
\begin{equation}
 \sqrt{-\gamma}\,\gamma_{ab}{\widehat
T}^{bc}=\lim_{r\to\infty}\sqrt{-h}\,h_{ab}T^{bc}\ .
\label{newstress}
\end{equation}

\section{The General Five-Dimensional Kerr--AdS Solution}

The five-dimensional Kerr-AdS$_5$ solution was first introduced by
Hawking, Hunter and Taylor-Robinson \cite{hawkingtwo}, where they
discussed its relevance to the AdS/CFT correspondence. In addition
to mass parameter $M$ and AdS radius $l$, this solution has two
rotation parameters $(a,b)$. The metric in Boyer-Lindquest-type
coordinates has the following form
\begin{eqnarray}
ds^2&=&-{\Delta_{r} \over{\rho}^2}
\left(dt-{a\sin^2{\theta}\over\Xi_a}d\phi-{b \cos^2{\theta}
\over\Xi_b}d\psi\right)^2+{\Delta_{\theta}\sin^2{\theta}\over\rho^2}\left(adt-{(r^2+a^2)
 \over {\Xi}_a} d\phi\right)^2\nonumber\\
& &+{(1+r^2/l^2)\over r^2\rho^2} \left(abdt-{b(r^2+a^2)
\sin^2\theta\over\Xi_a}d\phi-{a(r^2+b^2)
\cos^2\theta\over\Xi_b} d\psi\right)^2\nonumber\\
& &+{\rho^2\over\Delta_{r}}dr^2+{\Delta_{\theta}\cos^2{\theta}
\over\rho^2}\left(bdt-{(r^2+b^2)\over\Xi_b}d\psi\right)^2+{\rho^2\over\Delta_\theta}
d\theta^2\ ,\label{boyer}
\end{eqnarray}
where
\begin{eqnarray}
\rho&=&r^2+a^2\cos^2\theta+b^2\sin^2\theta,\nonumber\\
 \quad \Xi_a&=&1-a^2/l^2,
\quad \Xi_b=1-b^2/l^2\nonumber\\
\Delta_{r}&=&{1\over r^2}(r^2+a^2)(r^2+b^2)(1+r^2/l^2)-2MG, \nonumber\\
\Delta_{\theta}&=&1-a^2/l^2\cos^2\theta-b^2/l^2\sin^2\theta\
\label{m-funcs}.
\end{eqnarray}
The inverse temperature, computed by requiring regularity of the
Euclidean section, is given by:
\begin{equation} \beta={1 \over T}={2\pi
    r_{+}({r_{+}}^2+a^2)({r_{+}}^2
    +b^2)l^2\over 2r_{+}^6+r_{+}^4(l^2+b^2+a^2)-a^2b^2 l^2}\ \label{temp} .
\end{equation}
while the area of the horizon is
\begin{equation}
{\cal A}={2\pi^2 ({r_+}^2+a^2)({r_+}^2+b^2) \over {r_+}\Xi_a
\Xi_b}\ \label{area}.
\end{equation}
In these coordinates the angular velocities on the horizon have
the form:
\begin{equation}
{\Omega}^{a}_{H}= a {{\Xi}_a \over r_{+}^2+a^2},\quad
{\Omega}^{b}_{H}= b {{\Xi}_b \over r_{+}^2+b^2}\ .
\end{equation}
One of the features of the Kerr-AdS solution in Boyer-Lindquest
coordinates is the non-vanishing angular velocities
$\Omega^a_{\infty}=-a/l^2,\, \Omega^b_{\infty}=-b/l^2$, in the
$\phi$ and $\psi$ directions at spatial infinity. This is in
contrast to the asymptotically flat Kerr solutions case which has
a vanishing $\Omega_{\infty}$. It implies that observers at
spatial infinity associated with this coordinate system are not
co-rotating with the freely falling gas at infinity as in the
asymptotically flat Kerr case. Notice the dependence of the
angular velocities at infinity on the angular parameters of the
black holes. In principle, an observer or a gas at infinity can
have any angular velocity, it does not have to be related to the
rotation parameters of the black hole. We are going to realize the
importance of such a simple observation when we discuss the first
law.
\subsection{Previous Calculations}
In previous calculations \cite{adel,adel3} the counterterm method
has been used to calculate the action, stress tensor and conserved
charges of Kerr-AdS$_5$ (for a similar calculation but using
different time-like killing vector please see \cite{skenderis05}).
In this calculation the induced metric on the boundary is defined
as the hypersurface at $r\rightarrow \infty$, where $r$ is the
radial coordinate in the above Boyer-Lindquest-type form of the
Kerr-AdS$_5$ solution. Therefore, it was natural to choose the
boundary on which the dual field lives to be
\begin{equation}
ds^2= -dt^2+{2a\sin^2{\theta}\over\Xi_a}dtd\phi
+{2b\cos^2{\theta}\over\Xi_b}dtd\psi+l^2\left[{d\theta^2
  \over\Delta_{\theta}}
+{\sin^2{\theta} \over \Xi_a}d\phi^2+{\cos^2 \over \Xi_b}{\theta}d
{\psi}^2\right]\ .
\end{equation}
We are going to refere to this boundary as the rotating Einstein
universe (REU). Calculating the total energy and angular momentum
one obtains the following expressions:
\begin{eqnarray}
{\cal M}&=&{\pi l^2\over
96G\Xi_a\Xi_b}[7\Xi_a\Xi_b+{\Xi_a}^2+{\Xi_b}^2+72 G M/l^2]\ ,
\end{eqnarray}
and
\begin{equation}
{\cal J}_{a}={\pi M a \over 2 {\Xi_a}^2 \Xi_b},\quad   {\cal
J}_{b}={\pi M b \over 2 {\Xi_b}^2 \Xi_a}\ . \label{angular}
\end{equation}

The action is given by
\begin{eqnarray}
I_5&=&-{\pi \beta l^2 \over 96 \Xi_a\Xi_b
G}\left[12({r_+}^2/l^2)(1-\Xi_a-\Xi_b)+{\Xi_a}^2+{\Xi_b}^2
+\Xi_b\Xi_a\right.\nonumber\\
   & &\left.+12{r_+}^4/l^4-2(a^4+b^4)/l^4-12(a^2b^2/l^4)({r_+}^{2}l^{-2}-1/3)-12 \right] \ .
\end{eqnarray}
The above physical quantities satisfy the following thermodynamic
relation
\begin{eqnarray}
S=\beta\left( {\cal M}-\Omega^{a}_{H} {\cal J}_a+\Omega^{b}_{H}
{\cal J}_b\right)-I_5={{\cal A}\over 4G}\ , \label{thermo}
\end{eqnarray}
The general variation of the total energy expressions can not be
put in the form of the first law \be d{\cal M}\neq T
dS+\Omega^{a}_{H} d{\cal J}_a+\Omega^{b}_{H} d{\cal J}_b.\ee

\subsection{Kerr-AdS$_5$ Revisited}

Another natural conformal boundary for the Kerr-AdS solution is
the Einstein universe (EU)\be ds^2= -dT^2+l^2\, \left[
d\Theta^2+\sin^2{\Theta}d\Phi^2 +\cos^2{\Theta} d\psi^2\right].
\ee
\\ This metric is the hypersurface at $y\rightarrow \infty$ for
any asymptotically AdS solution in global coordinates, with $y$ as
a radial coordinate. Performing the following coordinate
transformations \cite{Hawkingone}\bea &&\Xi_a y^2\,
\sin^2{\Theta}=(r^2+a^2)\,\sin^2{\theta}\hspace{.4
in}\,\Phi=\phi+a\,t/l^2\hspace{.3 in}\,T=t \nonumber\\
&&\Xi_b y^2\, \cos^2{\Theta}=(r^2+b^2)\,\cos^2{\theta}\hspace{.4
in}\Psi=\psi+b\,t/l^2\ \label{xtransform}\eea the Kerr-AdS$_5$
solution (\ref{boyer}) take the following form, which is
manifestly asymptotic to AdS spacetime \cite{Gibbons}; \bea
ds^2=&-&(1+y^2/l^{2})dT^2+{ dy^2 \over 1+y^2/l^2-{2\, M \over
\Delta_{\Theta} y^2}}+y^2\,d\Omega_{3}^{2}\nonumber\\&+&{2M \over
{\Delta_{\Theta}}^3 y^2}(dT-a\, \sin^2{\Theta}\,d\Phi-b\,
\cos^2{\Theta}\,d\Psi)^2+ ...\ \label{non-rotating}\eea where \be
\Delta_{\Theta}=1-a^2/l^2\,\sin^2{\Theta}-b^2/l^2\,\cos^2{\Theta}\hspace{.5
in}\,
d\Omega_{3}^2=d\Theta^2+\sin^2{\Theta}\,d\Phi^2+\cos^2{\Theta}\,d\psi^2\ee

In this coordinate system and other coordinate systems used in
this paper, we are going to keep the thermodynamic quantities as a
function of the outer horizon radius, $r_{+}$, in the
Boyer-Lindquest-type coordinates in order to compare different
expressions obtained using different boundary metrics. Using
counterterms to calculate the action and total energy for
Kerr-AdS$_5$ in these coordinates with the Einstein universe as
our boundary metric, one gets the following
 \be {\cal M'}={\pi \over 32\,\Xi_a^2\,\Xi_b^2
}\left[MG(16\,\Xi_a+16\,\Xi_b-8\,\Xi_a\,\Xi_b)+{3\,{{\Xi_a}^2{\Xi_b}^2
\, l^2 \over G}}\right]. \label{energy} \ee  \be I_{5}'={\pi \beta
\over 32\,G \,l^2\, \Xi_a\Xi_b}
[4\,(r_{+}^{2}+a^{2})\,\,(r_{+}^{2}+a^{2})\,(l^2/r_{+}^2-1)+3\,\Xi_a\,\Xi_b\,l^4]\ee
The angular momenta are the same as in (\ref{angular}). The above
quantities satisfy the following thermodynamic relation \bea
S=\beta ({\cal M'}-\Omega {\cal J})-I_{5}'={{\cal A} \over 4 G}\ ,
\label{firstlaw} \eea  Also, they satisfy the first law
 \be d{\cal M'}= T\,dS+\Omega\,dJ.\ee
It is worth mentioning that the same coordinate system has been
considered in a background method calculation used by Gibbons,
Perry and Pope \cite{GPP1} to produce the action and other
physical quantities for Kerr-AdS$_D$. Their expressions satisfy
the above statistical relation (\ref{thermo}) and the first law
(\ref{1stlaw}). In a more recent work \cite{GPP3} the same authors
considered the vacuum energy of a Kerr-AdS$_5$ black hole and
argued that it is the same as that of AdS space (i.e., $E_c=
{3\pi\,l^2 \over 32 G}$ ). As we have discussed in section 2
counterterms can be used to obtain the same consistent results
produced by the background method. Furthermore, it produces the
correct quantities dual to the Casimir energy or the conformal
anomaly on the field theory side. As we have mentioned earlier,
the boundary field theory lives on Einstein Universe (22). Using
results of field theory on the Einstein universe (See for example
\cite{BD}), one can check that the Casimir energy and conformal
anomaly for the boundary field theory match that calculated using
the counterterm method. The Casimir energy is given by \be
E_{casimir}={3\,N^2 \over 16 l},\ee and the trace anomaly vanishes
on both sides \be T^{\mu}_{\mu}=0.\ee

\section{The First Law, Counter-terms and Kerr-AdS$_5$} We would like to discuss
the first law for Kerr-AdS$_5$ upon using the REU in (16) as a
boundary metric and write an expression for the variation of the
total energy in terms of the relevant thermodynamic parameters. It
is important to remind the reader that the simple form of the
first law is due to thermodynamic quantities that were measured by
an observer at rest relative to a free thermal gas at infinity.
For example, the total energy of the Schwarzschild or Kerr black
hole calculated using the ADM mass is the energy measured by an
observer at rest relative to the hole at infinity. Obviously, a
non-inertial observer measures different energy due to
non-inertial forces that might appear in his frame. Concerning
rotation, there are two types of non-inertial forces that appear
in a rotating frame; centripetal force and Coriolis force.
Coriolis force does not depend on the size of the system,
therefore, it would not contribute to the thermodynamic energy of
the system. As we have seen in the previous section, and as
pointed out in \cite{GPP1}, the variation of the total energy,
obtained using (16) as a boundary, can not be put in the form of
the first law. But, it can be written as
 \be d{\cal M}= T dS+\Omega^{a}_{H} d{\cal
J}_a+\Omega^{b}_{H} d{\cal J}_b+{\cal J}_a
d\Omega^{a}_{\infty}+{\cal J}_b d\Omega^{b}_{\infty}+d{\cal
M}_{cas}.\ee As one can see, the additional terms depend on the
$\Omega_{\infty}$'s variations, this is why the first law is
satisfied upon choosing EU as a boundary, since it has a vanishing
$\Omega_{\infty}=0$. Let us ignore the last term for a moment. The
energy not only depends on the usual extensive variables $(S,J)$,
but also on the intensive variables
$(\Omega^{a}_{\infty},\Omega^{b}_{\infty})$. This indicates that
this expression is not a well defined thermodynamic energy and we
better define another energy function which depend on extensive
variables only;\be {\cal M}={\cal
M'}+\Omega^{a}_{\infty}\,J^a+\Omega^{b}_{\infty}\,J^b ,\ee
therefore, \be d{\cal
M'}=TdS+(\Omega^{a}_H-\Omega^{a}_{\infty})\,dJ^a+(\Omega^{b}_H-\Omega^{b}_{\infty})\,dJ^b\ee
The meaning of this new energy function ${\cal M'}$ is simple, it
is the energy measured by an observer co-rotating with free gas of
particles at infinity. The time-like killing vectors of these two
energies are related by \be
\partial_{t'}=\partial_{t}+\Omega^{a}_{\infty} \partial_{\phi}+\Omega^{b}_{\infty}\partial_{\psi}\ee
As a result one has to use the time frame of the rotating free gas
at infinity to get a meaningful thermodynamic expression for the
energy of the system. This relation has been noticed in
\cite{GPP1,GPP2,skenderis05}, and we stress on its importance from
a thermodynamic point of view. As one can see $d{\cal M'}$ can be
put in the following form \cite{skenderis05} \be d{\cal
M'}=TdS+(\Omega^{a}_H-\Omega^{a}_{\infty})\,dJ^a+(\Omega^{b}_H-\Omega^{b}_{\infty})\,dJ^b+d{\cal
M}_c\ee where ${\cal M}_c={\cal M}_c(a,b)$ is the vacuum part of
the energy. Notice that $(a,b,r_{+})$ can be considered functions
of $({\cal J}_a,{\cal J}_b,S)$ regarding equation (14), and (18),
therefore, the last term violates the first law. The first law is
apparently violated because two independent physical quantities,
namely; $\Omega_H$ and $\Omega_{\infty}$, are related through
their dependence on the same parameter $a$. One should regard
$\Omega_{\infty}$ as a boundary property (i.e., of an observer, or
a gas at infinity) which does not have to dependent on the black
hole parameters. Notice that if $\Omega_{\infty}$ depends on
$r_{+}$ instead of $a$ the first law will be again violated.

\section{Another Coordinate System for Kerr-AdS$_5$}
In this section we present a different coordinate system for
Kerr-AdS$_5$ black hole with one rotation parameter\footnote{We
choose for simplicity, one rotation parameter, but it can be
easily generalized to two parameters and other dimensions as
well.}. This coordinate system can describe Kerr-AdS$_5$ from the
point of view of an observer rotating with respect to a freely
falling gas at infinity. The observer's angular velocity
$\Omega_{\infty}=c/l^2$ and that at the horizon are independent in
contrast to that of Boyer-Linquest-type coordinate. This
coordinate system can be obtained through the
 following coordinate transformation (\ref{non-rotating}) resulting in new coordinates $(t',r',\theta',\phi',\psi')$ \bea &&\Xi_c y^2\,
\sin^2{\Theta}=(r'^2+c^2)\,\sin^2{\theta'},\hspace{.4
in}\,\Phi=\phi'+c\,t'/l^2 \nonumber\\
&& y^2\, \cos^2{\Theta}=(r'^2)\,\cos^2{\theta'},\hspace{.3
in}\Psi=\psi',\hspace{.3 in}\,T=t'\ \eea Dropping the primes from
the new coordinates, leaves the metric component in the following
form \bea g_{tt}&=&-{r^2\over
l^2}-D_{\theta}+{2\,m\,\Xi_c\,\Delta_{\theta}^2 \over
r^2\,\Delta^3}+\,\,O({1 \over r^8})\nonumber\\
g_{t\phi}&=&{c\over l^2}\,{(r^2+c^2)\,\sin^2{\theta} \over
\Xi_c}-{2m\,a\,\sin^2{\theta}\, \Xi_c \Delta_{\theta} \over r^2\,\Delta^3}+\,\,O({1 \over r^8})\nonumber\\
g_{rr}&=&{l^2\over r^2}-{l^4 D_{\theta}\over
r^4}+2\,m\,l^4\,{\Xi_c^2 \over r^6 \Delta^2}+{l^6 \over r^6}\,[\,
D_{\theta}
+{b^4\over l^4}\,\sin^2{\theta}\,]+\,\,O({1 \over r^8})\nonumber\\
g_{\theta\theta}&=&{r^2+c^2\,\cos^2{\theta} \over
1-{c^2 \over l^2}\,\cos^2{\theta}}+\,\,O({1 \over r^8})\nonumber\\
g_{\phi\phi}&=&{(r^2+c^2)\,\sin^2{\theta} \over
\Xi_c}-{2m\,a^2\, \Xi_c \,\sin^4{\theta}\over  r^2\,\Delta^3}+\,\,O({1 \over r^8})\nonumber\\
g_{\psi\psi}&=&r^2\,\cos^2{\theta}, \eea where \be
D_{\theta}=1+c^2/l^2\, \sin^2{\theta}, \hspace{.5in}
\Delta_{\theta}=1-c^2/l^2\,\cos^2{\theta}-(a\,c)/l^2\,\sin^2{\theta}\ee
\be \Xi_c=1-c^2/l^2, \hspace{.3in} \Xi_a=1-a^2/l^2,
\hspace{.3in}\Delta=1-c^2/l^2\,\cos^2{\theta}-a^2/l^2\,\sin^2{\theta}\ee
The inverse temperature, $\beta$ and the area of the horizon
${\cal A}$ are the same as in (\ref{temp}) and (\ref{area}), but
the angular velocities at the horizon have the form:
\begin{equation}
{\hat{\Omega}}_{H}= a {(r_{+}^2/l^2+1) \over r_{+}^2+a^2}+c/l^2.
\end{equation}
Notice that, the previous two coordinate systems are special cases
of the coordinate system presented here, corresponding to $c=a$
and $c=0$. The hypersurface as $r\rightarrow \infty$ is chosen to
be our boundary metric in the counterterm calculation. The action
is given by \be \hat{I}_{5}={\pi \beta \over 96\,G \,l^2\, \Xi_a}
[12\,(r_{+}^{2}+a^{2})\,(l^2-r_{+}^2)+ {l^4\,\Xi_a (9\,
\Xi_c+c^4/l^4)\over \Xi_c}\,]\ee The energy associated with the
killing vector $\partial_t$ is given by \be {\hat{\cal M}}={\pi
\over 4\,\Xi_a^2\, }\left[\,M(3-a^2/l^2+2\,a\,c/l^2)\right]+{\pi\,
l^2\over 96\,\Xi_c\,G }\,(9\, \Xi_c+c^4/l^4). \label{energy2} \ee
Notice the dependence of the energy on $c$. The angular momentum
is \be {\cal J}={\pi M a \over 2 {\Xi_a}^2}.\ee All the above
quantities satisfy the statistical relation (28). The Casimir
energy and conformal anomaly of the boundary field theory
predicted from geometry side are given by \be E_c={\pi\,l^2 \over
96 \Xi_c}\,[9\,\Xi_c+ c^4/l^4]\ee and
\begin{eqnarray} {T_a}^a=-{c^2 N^2 \over 4 \pi^2
l^6}\left[c^2/l^2\cos^2\theta(3\cos^2\theta-2)-\cos 2
\theta\right]\ ,
\end{eqnarray}
which match exactly the expressions of the Casimir energy and
trace anomaly for $D=4$ $N=4$ SYM theory on the rotating Einstein
universe with angular velocity $\Omega_{\infty}=c/l^2$. Notice
here that the conformal invariance is broken because of the
non-vanishing angular velocity (i.e. $\Omega_{\infty}=c/l^2$) at
infinity not the black hole rotation parameter
$a$. \\
\subsection{First Law}
Now following the discussion on the previous section, the energy
associated with the killing vector is given by $\partial_t
+\Omega_{\infty}\,\partial_{\phi}$, \be {\cal M}={\pi \over
4\,\Xi_a^2\, }\left[\,M(3-a^2/l^2)\right]+{\pi\, l^2\over
96\,\Xi_c\,G }\,(9\, \Xi_c+c^4/l^4). \label{energy2} \ee The
energy and angular velocity \be{\Omega}_{H}=
{\hat{\Omega}}_{H}-\Omega_{\infty}=a {(r_{+}^2/l^2+1) \over
r_{+}^2+a^2}, \ee satisfy both (28) and the first law as well; \be
d{{\cal M}}=TdS+{\Omega}_H\,d{\cal J}.\ee This is true as long as
we think of $c$ as a fixed input parameter. $c$ can be thought as
a fixed parameter as a consequence of fixing the boundary metric.
This serves as a boundary condition on the metric in the AdS/CFT
set up, for more details please see \cite{skenderis05}. The first
law is directly satisfied in agrement with the general results of
\cite{skenderis05}.

It is worth mentioning that the energy ${\hat{\cal M}}$ associated
with the killing vector $\partial_t$ and the angular velocity
$\hat{\Omega}_{H}$ satisfy both (28) and the first law; \be
d\hat{\cal M}=TdS+\hat{\Omega}_H\,d{\cal J}.\ee

It is intriguing to notice that if one allows $c$ to vary, it will
lead to an additional term in the first law proportional to
${d{\cal M}_{cas} \over dA}$. Varying $c$ is the same as varying
the area $A$ of the spatial part of the boundary metric. It allows
the existence of external forces that act on the thermal gas at
infinity. This term can be interpreted as the work done by surface
tension, since the system has a curved boundary and the energy
depends on the boundary surface area $A$. From the boundary theory
point of view this surface tension is nothing but the Casimir
pressure, in addition to the usual conformal pressure ({\it i.e.},
which is proportional to 1/3 of the energy density). Calculating
the compressibility of the Casimir pressure, one find that it is
non-negative for $0\leq c \leq l$. Instead of writing the
expression for compressibility, which is rather long, we draw the
compressibility as a function of $c$ in Figure 1.
\begin{figure}
\begin{center}
\epsfig{file = 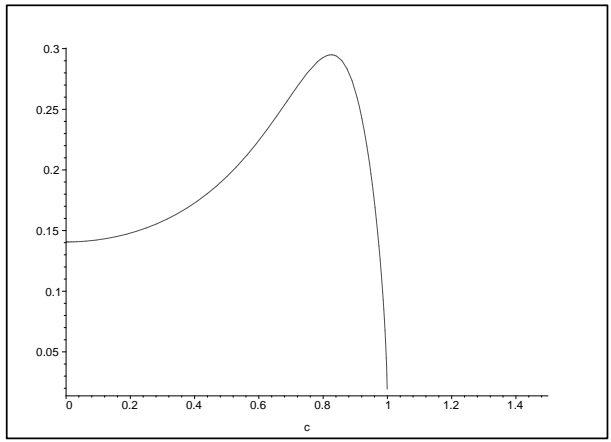, height=8cm,angle=-90,trim=0 0 0
0}\caption{ Compressibility K as a function of c, and l=1.}
\label{rootplot}
\end{center}
\end{figure}

This is a sign of thermodynamic stability of the system against
small changes in the volume of the the boundary. Notice here that
the range $0\leq c \leq l$ contains the values of $c$ that do not
change the metric signature and keep the velocity of any object
rotating with an angular velocity $\Omega_{\infty}$ less than that
of light.

\section{Concluding Remarks}
We use the standard counterterm method for the Kerr--AdS$_5$ case
to produce physical quantities that satisfy the first law of
thermodynamics. Here we choose the boundary metric to be the
non-rotating Einstein universe, similar to \cite{GPP1}. In this
work we point out the reason for the apparent violation of the
first law in some previous calculations \cite{adel,adel3}. We show
that the reason for the violation of the first law is not that REU
was chosen as the boundary metric but that it was rotating with an
angular velocity $\Omega_{\infty}=-a/l^2$ that depends on the
black hole rotation parameter, $a$. This boundary angular velocity
is that of an observer, or a thermal gas at infinity and does not
have to depend on the black hole parameters. Choosing to work with
a new coordinate system for Kerr--AdS$_5$ with arbitrary angular
velocity at infinity, one can show that the relevant physical
quantities satisfy the first law. This leads to the conclusion
that, in the counterterm method, angular velocities or other
quantities associated with a boundary metric should be independent
from the black hole parameters, otherwise, the first law might be
violated. It is interesting to notice that if we allow the angular
velocity to vary it will lead to an additional term in the first
law due to a surface tension on the boundary. The surface tension
is nothing but the Casimir pressure in the boundary theory. We
show that the existence of such pressure will not affect the
stability of the system since its compressibility is non-negative.

\section*{Acknowledgements}
The author would like to thank P. Argyres, S. Das, W. Merrell, A.
Shapere, and X. Wu for their comments and discussions which were
very helpful in different stages of this work. This work was
supported by Department of Energy Grant No. DE-FG01-00ER45832.

{}

\begin{thebibliography}{99999}
\bibitem{maldacena}J. Maldacena, Adv. Theor. Math. Phys. {\bf 2} 231
  (1998), hep-th/9711200.

\bibitem{witten}E. Witten, Adv. Theor.
  Math. Phys. {\bf 2} 253 (1998), hep-th/9802150.

\bibitem{gubklebpoly}S. S. Gubser, I. R. Klebanov and A.
M. Polyakov, Phys. Lett. {\bf B428} 105 (1998), hep-th/9802109.

\bibitem{hawkingtwo}S. W. Hawking, C. J. Hunter and M. M.
Taylor--Robinson, Phys. Rev. {\bf D59} (1999) 064005,
hep-th/9811056.
\bibitem{GPP1}G. Gibbons, M. Perry and C. Pope, Class.Quant.Grav. 22 (2005) 1503, hep-th/0408217.
\bibitem{GPP2}G. Gibbons, M. Perry and C. Pope, Phys. Rev. {\bf D} (2005) 084028, hep-th/0506233.
\bibitem{GPP3}G. Gibbons, M. Perry and C. Pope, Phys. Rev. Lett. 95 (2005) 231601, hep-th/0507034.
\bibitem{klemm2}M. Caldarelli, G. Cognola, and D. Klemm, hep-th/9908022.
\bibitem{skenderis05}I. Papadimitriou, K. Skenderis, JHEP 0508 (2005) 004, hep-th/0505190.
\bibitem{adel3} A. M. Awad and C. V. Johnson, Phys. Rev. {\bf D}63 (2001) 124023, hep-th/0008211.
\bibitem{mcnees}R. McNees, "A New Boundary Counterterm for Asymptotically AdS Spacetimes", hep-th/0512297.

\bibitem{klemm1}D.~Klemm, JHEP {\bf 9811}, 019 (1998), hep-th/9811126.


\bibitem{Hawkingthree}S.~W.~Hawking and H.~S.~Reall, Phys.\ Rev.\ {\bf
    D61}, 024014 (2000), hep-th/9908109.
\bibitem{Hawkingone}S.~W.~Hawking and G.~T.~Horowitz,
%``The Gravitational Hamiltonian, action, entropy and surface terms,''
  Class.\ Quant.\ Grav.\ {\bf 13}, 1487 (1996) gr-qc/9501014.
\bibitem{adel}A. M. Awad and C. V. Johnson, Phys. Rev. {\bf D61}
  (2000) 084025, hep-th/9910040.

\bibitem{ho}J.~Ho, hep-th/0005250.

\bibitem{lopez}K. Landsteiner and E. Lopez, JHEP {\bf 9912} 020 (1999),
hep-th/9911124.

\bibitem{horowitz} J. Bardeen and G. Horowitz, Phys. Rev. {\bf D60}
(1999) 104030, hep-th/9905099.




\bibitem{gubpeetkleb} S. S. Gubser, I. R. Klebanov and A. W. Peet,
Phys. Rev. {\bf D54} (1996) 3915, hep-th/9602065.


\bibitem{paper2} A. M. Awad and C. V. Johnson,
%{\it ``Scale Vs. conformal
% invariance in the AdS/CFT correspondence''},
hep-th/0006037.

\bibitem{Henningson}M. Henningson and K. Skenderis, J.H.E.P.  9807 023
  (1998), hep-th/9806087.

\bibitem{chalmers}G. Chalmers and K. Schalm, Phys. Rev. {bf D61} 046001
(2000), hep-th/9901144.

\bibitem{Balasubramanian}V.~Balasubramanian and
P.~Kraus, %``A stress tensor for anti-de Sitter gravity,''
  Commun.\ Math.\ Phys.\ {\bf 208}, 413 (1999), hep-th/9902121.


\bibitem{counterterm}R.~Emparan, C.~V.~Johnson and R.~C.~Myers,
%``Surface terms as counterterms in the AdS/CFT correspondence,''
  Phys.\ Rev.\ {\bf D60}, 104001 (1999), hep-th/9903238.




\bibitem{kraus}P.~Kraus, F.~Larsen and R.~Siebelink,
%``The gravitational action in asymptotically AdS and flat
spacetimes,''
  Nucl.\ Phys.\ {\bf B563}, 259 (1999), hep-th/9906127.

\bibitem{Brown}J. D. Brown and J. W. York, Phys. Rev. {\bf D47}, 1407
  (1993).

\bibitem{robstress}R.C.~Myers, hep-th/9903203.
\bibitem{robgary} G.~T.~Horowitz and R.~C.~Myers,
  Phys.\ Rev.\ {\bf D59}, 026005 (1999) hep-th/9808079.


%\bibitem{landau}L. D. Landau and E. M. Lifshitz, {\it'' Statistical Physics, Part 1''}, Pergamon (1980).
\bibitem{BD}N. D.  Birrell and P. C. W.
  Davies, {\it ``Quantum fields in curved space''}, Cambridge
  University Press (1982).  \bibitem{Gibbons}G.  W. Gibbons and S. W.
  Hawking, Phys. Rev {\bf D15}, 2752 (1977).



\end{thebibliography}
\end{document}